\documentclass[aps,twocolumn,superscriptaddress,showpacs,amssymb,prb,floatfix,preprintnumbers]{revtex4-1}

\usepackage{amsmath}
\usepackage{graphicx,color,ulem}
\usepackage[caption=false]{subfig}
\usepackage{epstopdf}

\usepackage[dvipdfm]{hyperref}

\hypersetup{colorlinks=true,linkcolor=blue,
  implicit=true,breaklinks=true,pagebackref=true,backref=true,
  bookmarks=true,bookmarksnumbered=true,hyperfootnotes=true,debug=true,
  naturalnames=false,citecolor=blue,pdfview=FitH,pdfstartview=FitH,hyperindex=true}

\begin{document}

% Use the \preprint command to place your local institutional report
% number in the upper righthand corner of the title page in preprint mode.
% Multiple \preprint commands are allowed.
% Use the 'preprintnumbers' class option to override journal defaults
% to display numbers if necessary
%\preprint{}

%Title of paper
\title{Anomalous nuclear magnetic resonance spectra in Bi$_2$Se$_3$ nanowires}

% repeat the \author .. \affiliation  etc. as needed
% \email, \thanks, \homepage, \altaffiliation all apply to the current
% author. Explanatory text should go in the []'s, actual e-mail
% address or url should go in the {}'s for \email and \homepage.
% Please use the appropriate macro foreach each type of information

% \affiliation command applies to all authors since the last
% \affiliation command. The \affiliation command should follow the
% other information
% \affiliation can be followed by \email, \homepage, \thanks as well.
\author{D. M. Nisson}
\author{A. P. Dioguardi}
%\author{P. Klavins}
\author{X. Peng}
\author{D. Yu}
\author{N. J. Curro}
%\email[]{Your e-mail address}
%\homepage[]{Your web page}
%\thanks{}
%\altaffiliation{}
\affiliation{Department of Physics, University of California, Davis, CA 95616, USA}

\date{\today}

\begin{abstract}
We report $^{209}$Bi nuclear magnetic resonance (NMR) spectra of in nanowires and single crystals of Bi$_2$Se$_3$.  Crystals were powdered to simulate the random orientation of the nanowires, and the spectra are compared with theoretical expectations derived via an orientational study of a single crystal.
The nanowire spectra are consistent with the randomly oriented powders, however we find an unusual suppression of signal intensity as a function of field orientation that may be associated with screening by surface currents.
%
%nanoparticles, powder, and a single crystal of Bi$_2$Se$_3$, and dependence of the spectrum of the single crystal as a function of angle. The ground powder spectra are taken before and after relief of mechanical strains, and before and after fixing in an epoxy resin in zero field to prevent alignment of grains in the applied magnetic field of 9 T. The nanoparticle and ground powder spectra all show an anomalous multi-peaked structure. The single-crystal angular dependent data indicate that this is the result of a new phenomenon outside of the sum of resonances in the individual powder grains. Summations of theoretical angular dependent spectra also fail to account for the powder spectrum, but the experimental angular dependence of the single crystal spectrum deviates significantly from theoretical expectations of NMR spectra of quadrupolar nuclei.
\end{abstract}

% insert suggested PACS numbers in braces on next line

\maketitle

\section{Introduction}

Topological insulators have attracted significant attention in the condensed matter community in recent years because of their potential applications in the field of spintronics.\cite{Pesin2012,Chen10072009,PhysRevLett.102.156603}
Their electronic band gap behaves as that of an insulator in the bulk, but because of large spin-orbit coupling
the topology of the bands are modified and protected surface states emerge which are gapless and conducting.
In these surface electronic states the electron spin is locked to the momentum, and thus the surface could carry spin currents.\cite{JPSJ.82.102001} In these materials the electronic backscattering on the surface is strongly suppressed, possibly making these currents dissipation-less and useful in spintronics applications.\cite{Moore2010}

The most actively studied experimental realization of topological surface states are materials in the bismuth chalcogenide family. Here we focus on Bi$_2$Se$_3$, the first such material that has attracted significant attention because of the discovery of topological surface states in 2009.\cite{Xia2009,PhysRevB.87.195202} Bi$_2$Se$_3$ has a rhombohedral crystal structure consisting of quintuple layers of triangular lattice planes, in the pattern Se-Bi-Se-Bi-Se.\cite{PhysRevB.87.195202} In principle this material should be a 350 meV band gap semiconductor in the bulk with conducting surface states carrying spin currents; however in practice there is an inherent chemical tendency for Se atoms to be missing from the lattice that significantly modify the bulk properties.\cite{hyde1974} These vacancies act as donors and render the bulk states degenerate, and often metallic, thereby making the surface transport properties difficult to observe.\cite{butch2010} Angle resolved photoemission spectroscopy (ARPES) studies have enabled detailed studies of the surface electronic dispersion\cite{RevModPhys.82.3045,Zhang2009,Bianchi2010}, but little is known about the microscopic behavior of the electronic states in these materials.\cite{PhysRevB.85.155301} We have studied the interactions between the electronic states and the nuclei through nuclear magnetic resonance (NMR).

Our previous study of the NMR properties of Bi$_2$Se$_3$ single crystals revealed a significant hyperfine coupling between the $^{209}$Bi nuclei and the bulk electrons.\cite{PhysRevB.87.195202} In principle, the hyperfine interactions between nuclei and electrons can give rise to spin-spin scattering and hence lead to the decoherence of polarized electron spins.\cite{PhysRevB.86.045451} Since the potential applications of topological insulators involve taking advantage of the spin polarization of currents, it is important to understand how the hyperfine interaction will affect the spins of the surface states. It is vital, therefore, to investigate the NMR response of surface nuclei in topological insulators.\cite{PhysRevLett.110.026602}  However, the lattice distances between the individual atoms in the triangular lattice planes of Bi$_2$Se$_3$ is about 4.14 \AA\, whereas the distance between the Bi monolayers is 4.3 \AA on average.\cite{LatticeConstants1998} This means that there were more than 1 million bulk nuclei for each surface nucleus in the large single crystals used in our previous work, which were on the order of 5 mm x 5 mm x 0.5 mm. Having enough surface nuclei to detect them against a bulk background requires a powder consisting of small particles with large surface-to-volume ratios. Even for samples ground via a mortar-and-pestle, the particle sizes are usually 10 $\mathrm{\mu}$m or larger in diameter, there are on the order of 30 000 or more bulk nuclei for each near-surface nucleus of $^{209}$Bi. Therefore it is difficult to discern the surface states against the bulk background in a magnetic resonance experiment in a random powder, unless special techniques are used to grow nanoscale particles.\cite{PhysRevLett.110.026602}

\begin{figure}%
\subfloat[][]{
  \includegraphics[width=\linewidth]{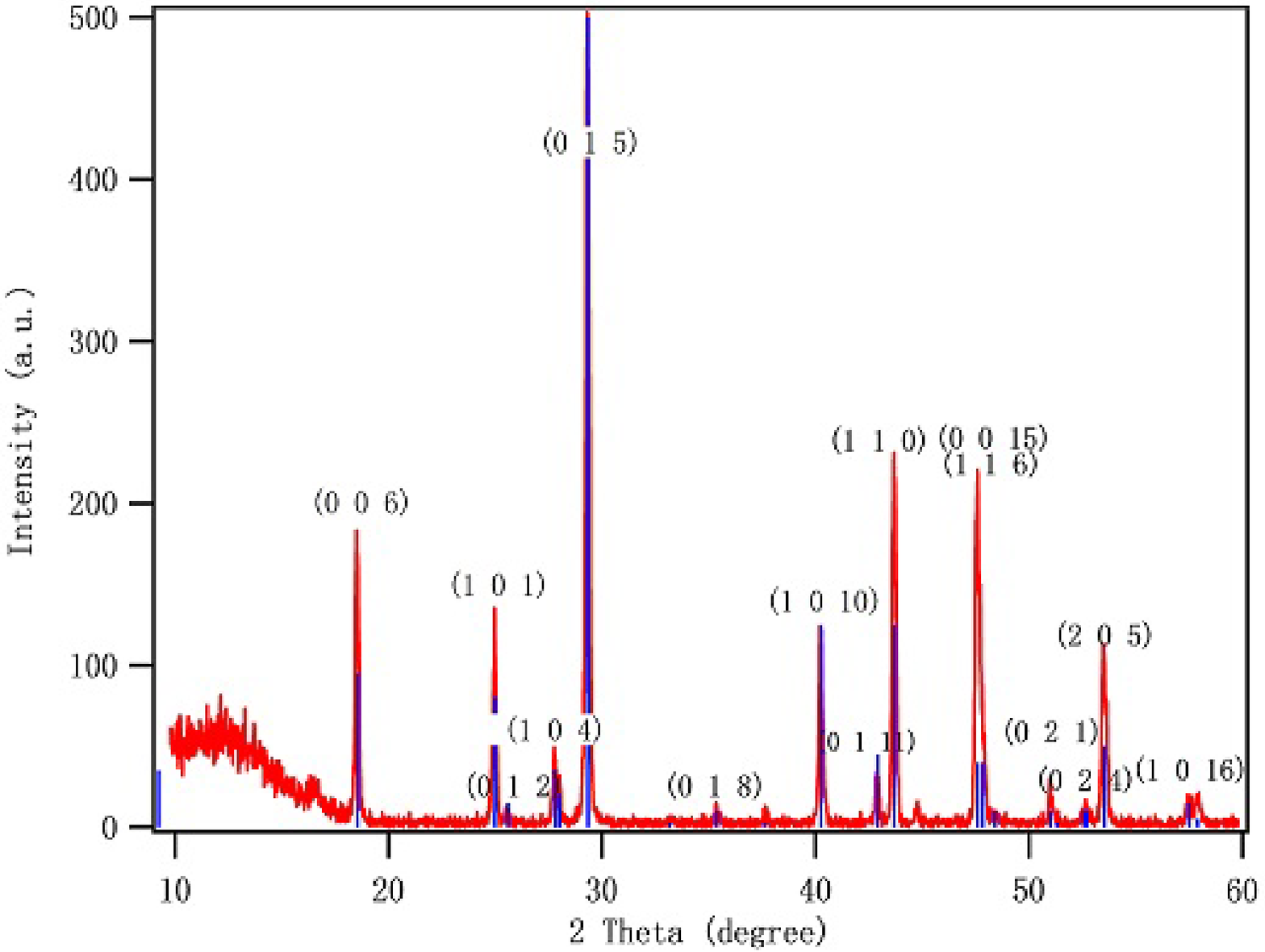}%
  \label{nano_xrd}}\\
\subfloat[][]{
  \includegraphics[width=0.4\linewidth]{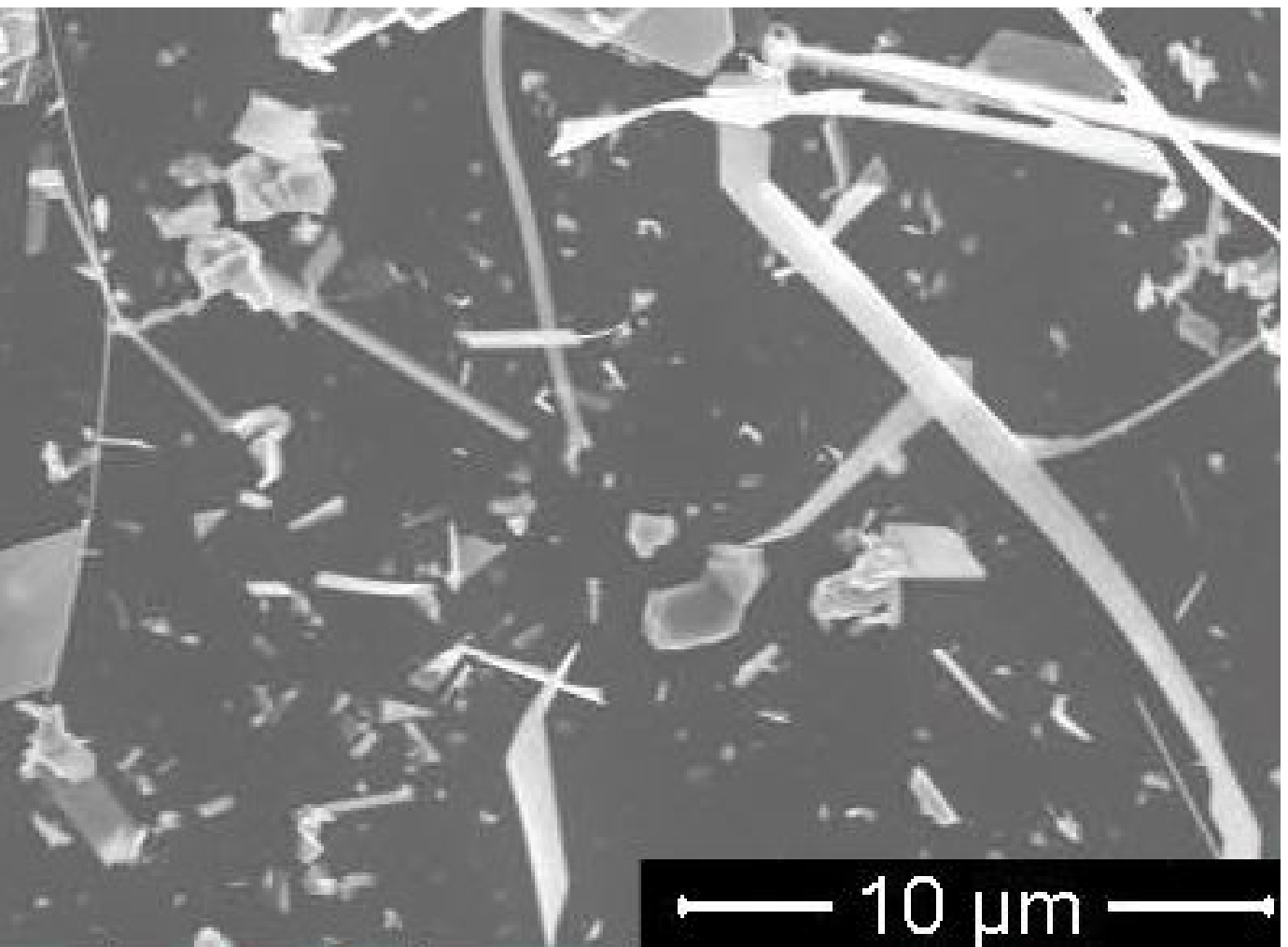}%
  \label{nano_sem10}}
\qquad
\subfloat[][]{
  \includegraphics[width=0.4\linewidth]{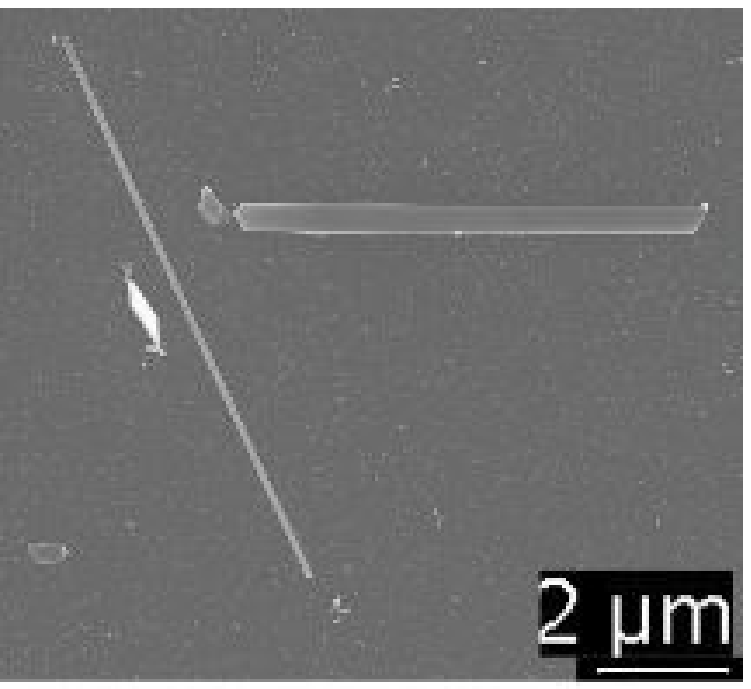}%
  \label{nano_sem2}}
  \caption{\protect\subref{nano_xrd} X-ray diffraction spectrum (red) of nanoribbons grown by the methods reported here. The phase purity of the nanowires is confirmed by the matching of peak locations between experimental data and predictions (blue vertical lines). \protect\subref{nano_sem10},  \protect\subref{nano_sem2} Scanning electron microscopy images of nanoribbons and nanoplates. }%
\end{figure}

Here we report a $^{209}$Bi NMR study in both Bi$_2$Se$_3$ nanowires and ground crystalline powders. A previous $^{125}$Te  NMR study of the related material Bi$_2$Te$_3$ revealed features in the powder spectra consistent with surface nuclei.\cite{PhysRevLett.110.026602} However, in order to discern how the hyperfine interaction may play a role in affecting the surface states, it is important to connect the behavior in single crystals with that in random powders. We have investigated the $^{209}$Bi NMR properties of the bulk states in micron scale powders of Bi$_2$Se$_3$. As particle size decreases, surface-to-volume ratio increases as a natural consequence of the fact that surface area increases in proportion to two length scales of the particle, while volume increases in proportion to three. The smallest particles used in this study are predominantly rod-shaped and plate-shaped. The surface area to volume ratio of a cylinder of radius $r$ and height $h$ is given by:
\begin{equation}
\frac{S}{V} = \frac{2 \pi rh + 2 \pi r^2}{\pi r^2 h} = \frac{2}{r} + \frac{2}{h}.
\end{equation}
The surface of a nanowire of Bi$_2$Se$_3$ has on the order of $n_S$ = 6 $^{209}$Bi nuclei nm$^{-2}$ while the volume has $n_V$ = 7.9 $^{209}$Bi nm$^{-3}$. Thus the ratio of surface to volume nuclei is
\begin{equation}
\frac{N_S}{N_V} = \frac{n_S S}{n_V V} = \frac{n_S}{n_V}\left(\frac{2}{r} + \frac{2}{h}\right).
\end{equation}
Since the nanowires used in our study have radii of the order of 100 nm, around 2\% of the nuclei in the nano particle sample should be the Bi nuclei near the surface, the ones with the strongest hyperfine interactions with the chiral electronic states. In contrast, for our powders ground from large single crystals with a mortar and pestle we expect on the order of 0.001\% of the nuclei reflect properties of the surface. Nevertheless, we find that both the nanowire and powdered samples both exhibit similar spectra, and we compare these results to the angular dependence of the spectrum of a single crystal. Surprisingly, however, the spectral intensity drops significantly as the field is rotated out of plane. This suppression of intensity does not appear to arise due to variations of either the spin-lattice relation or spin-spin decoherence rate as a function of angle.  We speculate, therefore, that this angular-dependent suppression reflects an anomalous screening of the radiofrequency NMR excitation pulses.

%where $\nu_{cc}$ is the quadrupolar splitting with $\overrightarrow{H} \parallel c$ and $\theta$ is the angle between $\overrightarrow{H}$ and $c$.
%Single-crystal spectra, as shown in Fig. \ref{scry_angdep}, have clearly resolvable transitions . The splitting goes to zero at the "magic angle", $\theta = 54.736^{\circ}$. At this angle, all of the transitions are, to first order, expected to occur at the same frequency, resulting in a much greater spectral intensity at the center frequency. However, a powder sample is a sum of many randomly-orientated single crystals, and so the resulting spectra will be an almost-continuous superposition of these nine-peak spectra with different splittings at different orientations.\cite{Samoson198529} Because the $^{209}$Bi nucleus has nine transitions, the powder spectrum should be a sum of many spectra with many different electric field gradient splittings. Therefore, we aim to compare experimentally observed $^{209}$Bi spectra with theoretical calculations. What we find is an anomalous result that has yet to be explained.

\section{Experimental Methods}

Single crystals of Bi$_2$Se$_3$ were grown by the Bridgman method described previously\cite{PhysRevB.87.195202} using stoichiometric amounts of Bi and Se in evacuated quartz ampoules. Powder X-ray diffraction using a Siemens D500 X-ray diffractometer confirms that the sample is phase-pure Bi$_2$Se$_3$.\cite{PhysRevB.87.195202}  Nanoribbons and nanoplates were grown from 99.999\% pure powdered Bi$_2$Se$_3$ provided by Alfa Aesar, Inc., by a chemical vapor deposition method modified from previous literature.\cite{Kong2010} The powder was heated to about $580\,^{\circ}\mathrm{C}$ in a tube furnace and nitrogen gas was flowed to carry the vapor to a Si substrate covered with 20 nm diameter Au nanoparticles from Ted Pella, Inc. The pressure was kept at about 0.6 Torr using a vacuum pump. The resulting nanoribbons were typically $40\,\mathrm{nm} \times 100\,\mathrm{nm} \times 30\,\mu\mathrm{m}$ and the nanoplates ranged from $5\,\mathrm{\mu m}$ to $50\,\mathrm{\mu m}$. X-ray diffraction confirms that nanoribbons and nanoplates grown by this method are phase-pure Bi$_2$Se$_3$ (Fig.~\ref{nano_xrd}).

\begin{figure*}[!]%
\subfloat[Experimental data]{
  \includegraphics[width=0.4\linewidth]{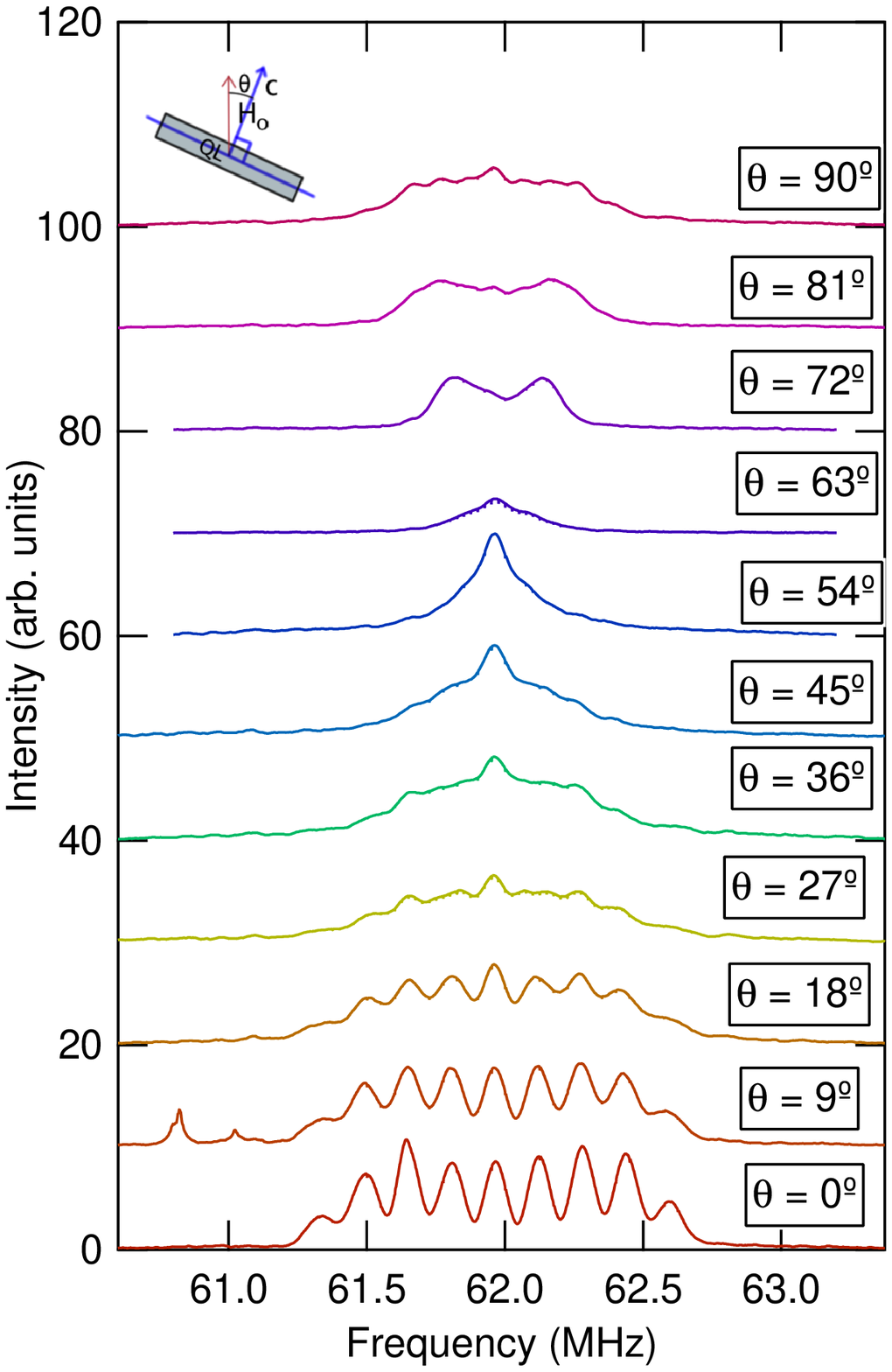}
  \label{scry_angdep_exp}
   }%
\qquad
\subfloat[Theoretical prediction]{
  \includegraphics[width=0.4\linewidth]{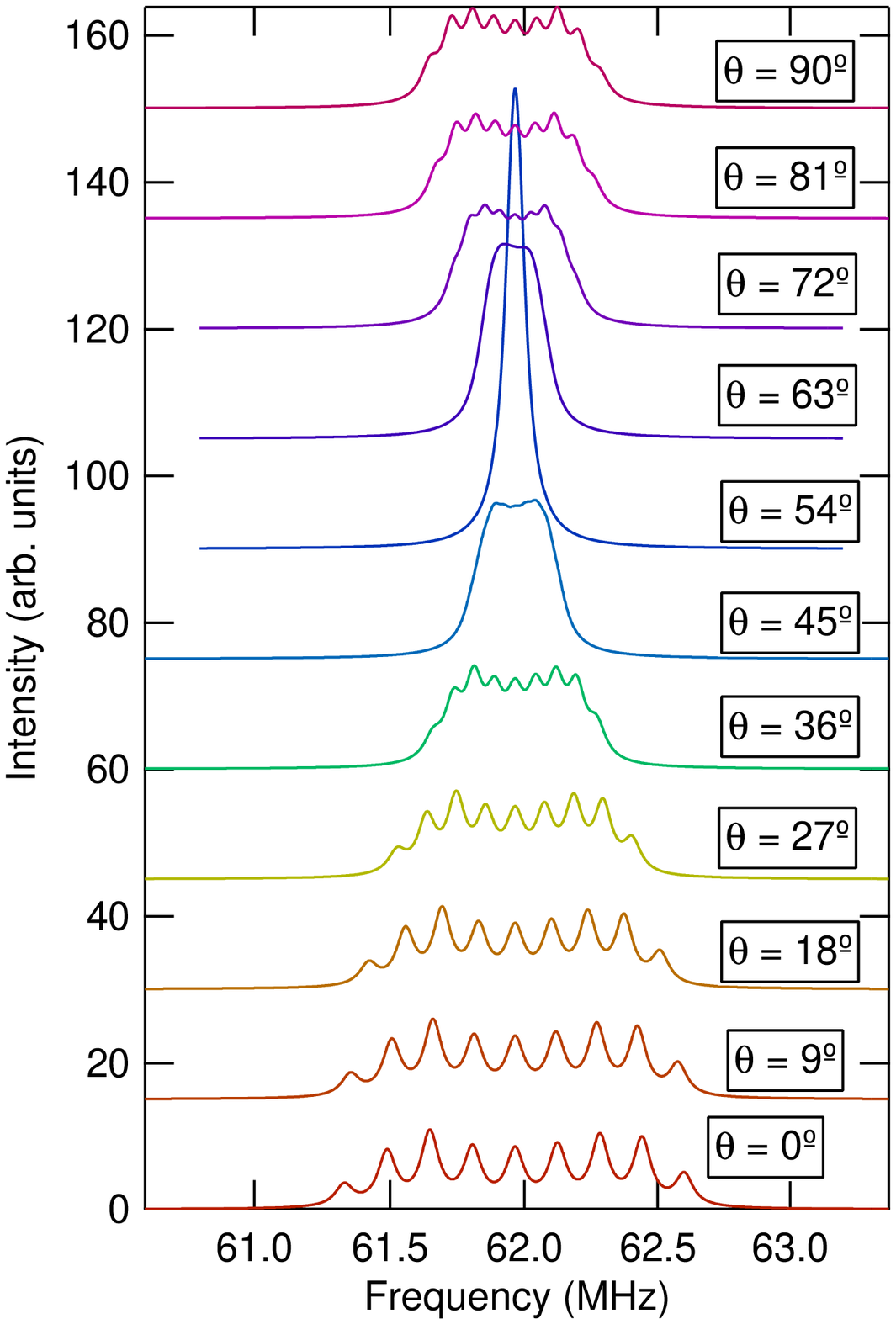}
  \label{scry_angdep_theo}
   }%
\caption{\label{scry_angdep}\protect\subref{scry_angdep_exp} Spectra of Bi$_2$Se$_3$ single crystal taken as a function of the angle $\theta$ between the magnetic field and the normal to the quintuple layers (inset), at 10 K and 9 T. The single crystal spectra at nonzero angles reveal intensity reductions  as shown also in Fig.~\ref{scry_angdep_area}.  \protect\subref{scry_angdep_theo} Theoretical angular-dependent spectra of Bi$_2$Se$_3$ predicted using the amplitudes from the spectrum at $\theta = 0$ and the formula $\Delta \nu_Q = \frac{1}{2} \nu_Q \left(3\cos^2 \theta - 1\right)$, showing the expected  increase in center amplitude near $60^{\circ}$.}
\end{figure*}

NMR measurements were performed on powder samples at a magnetic field of 9 T and a temperature of 10 K using a Quantum Design Physical Properties Measurement System (PPMS) for field and temperature control. The nuclear spin of the Bi nucleus in this material is described the the Hamiltonian:
\begin{equation}
\mathcal{H} = \gamma H_0 \hat{I_z} + \frac{h \nu_{cc}}{6}\left[3\hat{I_c}^2 - \hat{I}^2\right] + \mathcal{H_\mathrm{hf}},
\end{equation}
where
\begin{equation}
\hat{I_c} = \hat{I_z} \cos \theta + \hat{I_x} \sin \theta,
\end{equation}
$\gamma = 0.6842$ kHz/G is the gyromagnetic ratio, $H_0$ is the external field of 9 T for mortar-and-pestle powders or 8.795 T for nanowires, $\hat{I}_\alpha$ are the nuclear spin operators, $\nu_{cc}$ is the electric field gradient along the $c$-direction, $\theta$ is the angle between the magnetic field and $c$-axis, and $\mathcal{H}_{\mathrm{hf}}$ is the hyperfine interaction between the nucleus and the electronic spins.\cite{CPSbook} In this case there are nine transitions equally split by an amount $\Delta \nu_Q$ given by:
\begin{equation}
\Delta \nu_Q(\theta) = \frac{\nu_{cc}}{2}\left(3 \cos^2 \theta - 1\right).
\end{equation}
For the single crystal a single-axis goniometer probe was used to take spectra and relaxation rate measurements as a function of angle, $\theta$, in $9^{\circ}$ intervals. Single-crystal spectra are shown in Fig. \ref{scry_angdep_exp}. NMR measurements were conducted on the nanoribbons  by filling a gelatin capsule with approximately 10 mg of material and placing the capsule within the NMR coil.  Data were acquired at 8.795 T and 5 K, and spectra are shown in Fig. \ref{powder_exp_nano}. In order to compare the nanoribbons with the material grown via bulk synthesis, crystals were ground to a powder using a mortar and pestle.  These samples were first measured as-prepared, then mixed with powdered Si and annealed in an Ar atmosphere at $327\,^{\circ}\mathrm{C}$ for 135 minutes to relieve mechanical strains.  To rule out possible alignment of the powder grains with the 9 T magnetic field, we then fixed the powder in an epoxy resin at zero magnetic field then measured the powder sample again. As seen in Fig. \ref{powder_exp}, the spectra remain unchanged indicating that neither strain-induced broadening or partial alignment of the random powder are significant effects. By integrating the spectra of the mortar-and-pestle and nanowire samples we estimate their respective Knight shifts, $K$, to be 0.58(5)\% and 0.62(5)\%, respectively. The Knight shift in Bi$_2$Se$_3$ is known to be correlated with the density of charge carriers in the material.\cite{PhysRevB.87.195202} From the previously obtained relationship the carrier concentrations of the ground powder and nanowires are respectively $n_\mathrm{ground} \approx 9.9 \times 10^{-18}\,\mathrm{cm}^{-3}$ and $n_\mathrm{nano} \approx 1.3 \times 10^{-19}\,\mathrm{cm}^{-3}$, which attests to the high quality of the nanoribbon samples.

\section{Results and Discussion}

\begin{figure}
\includegraphics[width=\linewidth]{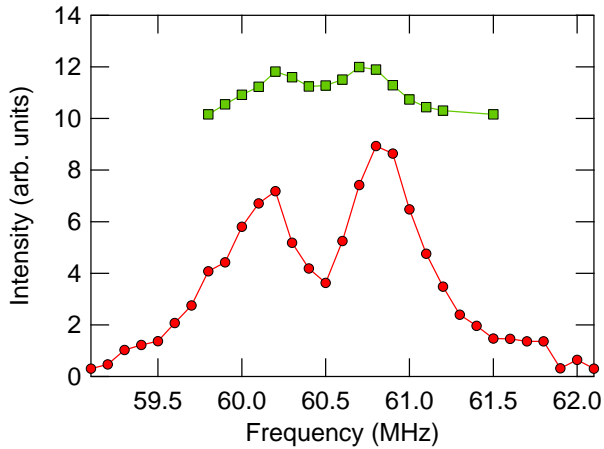}%
  \caption{\label{powder_exp_nano} NMR spectra of the Bi$_2$Se$_3$ nanoparticles at 5 K (red circles) and 60 K (green squares).}
\end{figure}

\begin{figure}%
  \includegraphics[width=\linewidth]{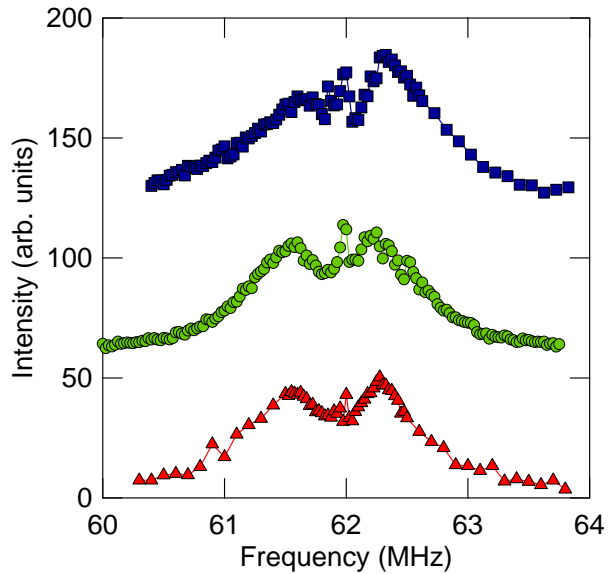}%
  \caption{\label{powder_exp}Experimental 10 K, 9 T $^{209}$Bi spectra of Bi$_2$Se$_3$ ground with a mortar and pestle, showing the spectrum of bulk states. The data was first taken with the powder as-made (blue squares), then after strain relief (green circles), then after fixing in epoxy resin in zero field to prevent alignment of grains (red triangles). All three spectra are dominated by two broad peaks, as in the nanoparticle case (Fig. \ref{powder_exp_nano}).}%
\end{figure}

As seen in Fig. \ref{scry_angdep_exp}, the single crystal NMR spectra vary strongly with field orientation.  In a  powder, each grain or crystallite will have a random orientation with respect to the field, giving rise to the broad powder patterns observed in Figs. \ref{powder_exp_nano} and \ref{powder_exp}.  In order to quantify these spectra, we measured the detailed angular dependence of the single crystal.  The observed spectra agree well with the predicted angular variation of the quadrupolar interaction (Fig. \ref{scry_angdep_theo}), and  reveal a quadrupolar splitting $\nu_{cc} = 160$ kHz, which is comparable to the line width in most samples.\cite{PhysRevB.86.075137, PhysRevB.87.195202} As a result, the powder patterns are not expected to display the sharp horns often observed in other NMR powder spectra of quadrupolar nuclei.\cite{PhysRevB.49.16321}  However, we find that the observed spectra differ significantly from the theoretical expectations.

\begin{figure}
  \includegraphics[width=\linewidth]{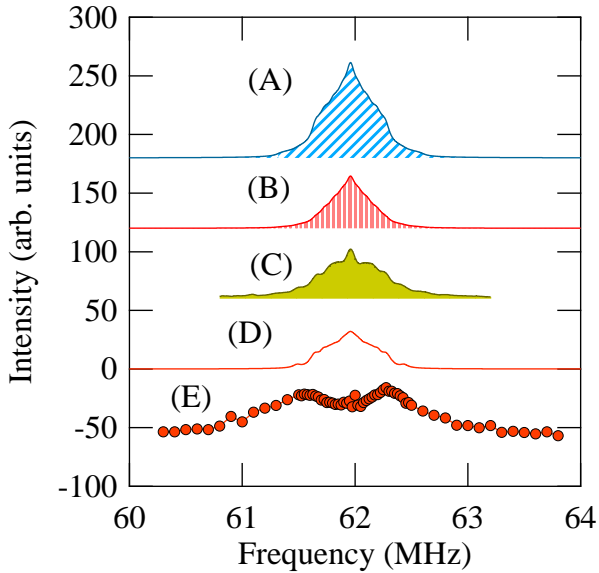}%
  \caption{\label{powder_theo} $^{209}$Bi powder spectra simulated by three separate methods, (A), (B), and (C), as described in the text. (D) Sum of theoretical amplitude spectra with the same number of angle points as the experimental data; this spectrum exhibits bumpiness similar to the solid gold pattern. (E) Experimental spectrum of mortar-and-pestle ground sample.}
\end{figure}

Figure~\ref{powder_theo} shows theoretical powder patterns for the Bi calculated via three separate methods. In method (A) the spectrum was calculated via a weighted sum of spectra from each of the nine satellite transitions.  Each transition is given by a Lorentzian: $f_n(\omega,\theta) = \frac{A_n}{\pi} \frac{\Delta \omega_n}{\left(\omega - \omega_n(\theta)\right)^2 + \Delta \omega_n^2}$, where  $\Delta \omega_n$ is the line width of the transition $n$ (set to a fixed value of 40 kHz based on the single crystal spectra), $\omega_n(\theta) = \omega_0 + n\Delta \nu_Q (\theta)$, $\omega_0 = 62$ MHz is the center frequency, and $A_n$ is the amplitude of the $n^{th}$ transition: $A_n = \left[I(I + 1) - \left(n - \frac{1}{2}\right)\left(n + \frac{1}{2}\right)\right]e^{-2\tau / T_{2,n}}$.  The spin-spin relaxation rates, $T_{2,n}$ can be estimated via Redfield theory, and are give by:\cite{CPSbook,NickCharlieEFK}
\begin{equation}
\frac{1}{T_{2,n}} = \frac{1}{2T_1} \left\{
\begin{array}{cc}
17 + r & n = \pm4 \\
31 + r & n = \pm3 \\
41 + r & n = \pm2 \\
47 + r & n = \pm1 \\
49 + r & n = 0
\end{array}
\right.
\end{equation}
where $r = \frac{{H}_{\mathrm{hf},H_0 \parallel c}}{{H}_{\mathrm{hf},H_0 \perp c}}$ is the anisotropy ratio of the hyperfine field at the Bi sites.  Here we have taken $r=1$ for concreteness.  The powder pattern is then given by the sum of all of these transitions at all angles:
\begin{equation}
\label{lor_sum_int}
S(\omega) = 2\pi \sum_{n = -4}^{+4}  \int\limits_0^\pi f_n(\omega,\theta) \sin \theta\,\mathrm{d}\theta.
\end{equation}
This approach yields a spectrum that is peaked close to $\omega_0$, as seen in Fig. \ref{powder_theo}, but does not agree with the experimental result (E).

In order to discern the origin of the discrepancy with the experimentally observed powder spectrum, we also computed the spectrum using the amplitudes $A_n$ as measured in the single crystal (Fig. \ref{scry_angdep_exp}) for $\theta = 0^{\circ}$, rather than the theoretical amplitudes discussed above.  Although the satellite amplitude typically decreases with frequency distance from center\cite{BiFeO3NMR,PhysRevB.49.16321},  the Bi satellite amplitudes in Bi$_2$Se$_3$ have been observed to differ from theoretical expectations, and have been attributed to $T_2$ effects.\cite{PhysRevB.86.075137}  The theoretical single-crystal spectra using these fitted amplitudes are shown in Fig. \ref{scry_angdep_theo}, and the resulting powder pattern is shown as  spectrum (B) in Fig. \ref{powder_theo}.  Again, the powder pattern  still exhibits a large peak at $\omega_0$ and disagrees with experiment.

In method (C) the theoretical powder pattern was determined simply by summing the measured single-crystal spectra as a function of angle, weighted by $\sin \theta$. In this case, the pattern reveals several bumps not present in the first two; however these bumps most likely arise because of the limited set of angles used (only eleven experimental spectra).  To test this idea, we computed the theoretical spectrum using the same limited set of angles, which reproduces the bumps due to the relatively large intervals between angles (Figure~\ref{powder_theo}D).

It is clear that the theoretical powder spectra (A-D) in Fig. \ref{powder_theo} are inconsistent with the observed spectrum (E) and those in Fig. \ref{powder_exp_nano}. Whereas the theoretical powder pattern indicates a large spectral weight at the center, the powder and nanoparticle spectra show suppressed intensity at the center. All of these powder spectra taken from the same batch of Bi$_2$Se$_3$ crystals reveal  two broad peaks with a small, narrow central peak regardless of whether strains are relieved or crystals are prevented from aligning in the field by epoxy resin. The spectrum for the nanoparticles of Bi$_2$Se$_3$ reveals a similar anomalous structure. There are two broad peaks at roughly the same frequencies as the powder spectra, however the narrow central peak is absent in the nanoparticle spectrum. Previous studies on Bi$_2$Se$_3$ grown by other methods\cite{PhysRevB.86.075137} do not reproduce the narrow central peak that we see in our single-crystal spectra at higher angles, which may be due to a small impurity phase or some other feature unique to our samples.

\begin{figure}%
	\includegraphics[width=\linewidth]{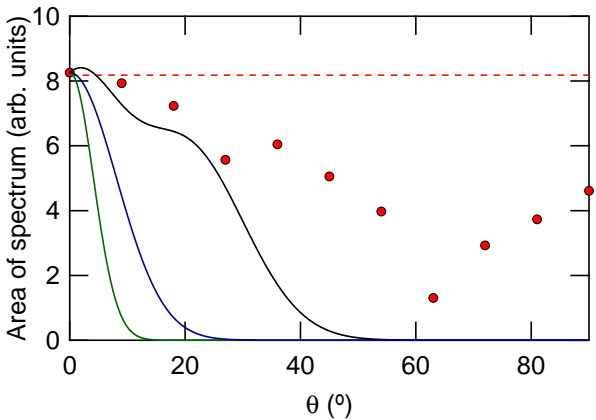}
	\caption{\label{scry_angdep_area} Total areas under spectrum as a function of angle: experimental (red circles), theoretical expectation (dotted line), and attempts to reproduce powder spectrum as described in text (black, blue, and green lines).}
\end{figure}

\begin{figure}%
   \includegraphics[width=\linewidth]{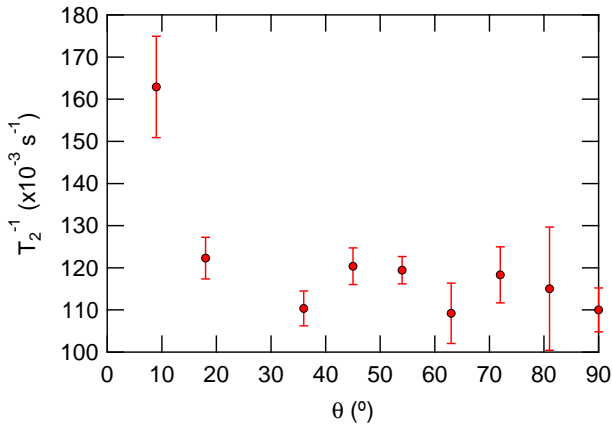}%
   \caption{\label{t2_vs_angle} Spin-spin relaxation rate as a function of angle for single crystal of Bi$_2$Se$_3$. The relaxation times are independent of angle above 18$^{\circ}$, thus ruling out spin-spin decoherence as a cause for the anisotropy of signal intensity.}
\end{figure}

The discrepancy between the theoretical and observed powder patterns can be explained if the spectral intensity is significantly suppressed for large angles $\theta$.  This discrepancy is even evident in the single crystal spectra shown in Fig. \ref{scry_angdep}. The overall shape of the experimental spectrum follows expectations as $\theta$ approaches $90^{\circ}$, but there appears to be a significant drop in intensity, particularly near 60$^{\circ}$.  The total spectral area is shown in Figure~\ref{scry_angdep_area}, which indicates that roughly 50\% of the intensity is missing for $\theta\gtrsim 45^{\circ}$. The area under the spectra should remain constant, reflecting the number of nuclei contributing to the NMR signal. Such an effect could be caused by a strong angular dependence of $T_2$.  However, direct measurements of $T_2$ in the single crystal as a function of $\theta$ do not reveal any anomalous behavior (see Fig. ~\ref{t2_vs_angle}).  The origin of this behavior is not understood, but could be related to a suppression of the rf screening at this angle.

\begin{figure}%
	\includegraphics[width=\linewidth]{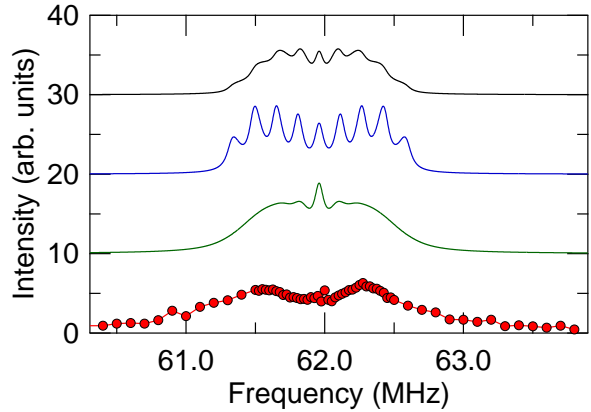}
	\caption{\label{powder_simspec} Simulated powder spectra (black, blue, and green lines) and observed spectrum in the mortar-and-pestle sample (red circles) using spectral weighting functions $\beta(\theta)$ as described in the text.}
\end{figure}

For the powder pattern, we also find that the spectra for higher angles contribute less to the overall intensity.  To see this, we computed the powder pattern by introducing a new parameter, $\beta(\theta)$ representing a weighting factor as a function of angle:
\begin{equation}
\label{lor_sum_int2}
S(\omega) = 2\pi \sum_{n = -4}^{+4}  \int\limits_0^\pi \beta(\theta) f_n(\omega,\theta) \sin \theta\,\mathrm{d}\theta.
\end{equation}
Using the approach (B) now with the additional $\beta(\theta)$ factor in Eq.~\ref{lor_sum_int2}, we are able to produce a function that resembles the observed two-peak powder pattern in shape but not in frequency width using a function $\beta(\theta) = e^{-\theta^2/(\Delta \theta_0)^2} + \alpha e^{-(\theta - \theta_1)^2/(\Delta \theta_1)^2}$, where $\Delta \theta_0 = 0.2, \Delta \theta_1 = 0.25, \theta_1 = 0.35, \alpha = 0.8$ (Fig. \ref{powder_simspec}, black line). An attempt with a single Gaussian and a smaller range of angles, corresponding to the above parameters except for $\alpha = 0$, reproduces the width but leaves the nine-peaked structure (Fig. \ref{powder_simspec}, blue line). The respective $\beta(\theta)$ for the two parameter sets are the black and blue lines in Fig. \ref{scry_angdep_area}.  The $\beta(\theta)$ weighting factor clearly drops off dramatically for $\theta \gtrsim 10^{\circ}$.  This result suggests that only grains near $H \parallel c$ contribute to the signal in the powder samples.  Note that there is an apparent minimum close to the magic angle of $54.7^{\circ}$, which may be related to the fact that multiple satellite transitions are excited at this angle. 

Although the powder spectrum appears to arise solely from grains near $H \parallel c$, the value of $\nu_Q$ in that regime changes little as a function of angle. As a result, if there were no additional line-broadening mechanisms in the powder samples, the spectrum would be expected to resemble the blue line in Fig. \ref{powder_simspec}, where all nine transitions are visible. If the hypothesis of suppressed crystal orientations is true, then the broad spectra that we observe could be explained by an increase in the electric quadrupolar broadening of line widths in the powder samples compared to single crystals. This is consistent with the fact that, even when relieved of mechanical strains, the powder samples still produce spectral "tails" at the low and high frequency ends that are absent from the single-crystal spectra. Perfoming the aforementioned calculations using parameters $\Delta \theta_0 = 0.1, \Delta \theta_1 = 0.1, \theta_1 = 0.35, \alpha = 0.2$ for $\beta(\theta)$ (Fig. \ref{scry_angdep_area}, green line), with an additional quadrupolar broadening $\nu_Q = 80\,\mathrm{kHz}$, produces a spectrum similar to what we observe, but with a much taller central peak, compared to the surrounding resonances, than in the experimental spectra (Fig. \ref{powder_simspec}, green line).  Strain-induced broadening of the EFG distribution would also explain the extended tails observed in the powder spectra. 

In order to test whether this strong angular suppression is a result of spin-spin decoherence, we measured the spin-echo decoherence rate, $T_2^{-1}$ as a function of angle in the single crystal, shown in Fig. \ref{t2_vs_angle}. There is no clear evidence for any significant angular dependence in the single crystal, but we cannot rule out that  $T_2^{-1}$ may exhibit a stronger angular dependence  in the powder or nanoparticle samples.

Another potential explanation for the suppression of signal in the powder/nanoparticle materials for large angles is that the radiofrequency field $H_1$ is altered by the presence of surface currents in these topological insulator samples.  In order to excite the nuclei in the material, the NMR coil creates a dynamic field $B_1$, which gives rise to a field $H_1 = B_1/\mu_0 - M$ inside the material of interest. For conductors, $H_1$ becomes non-uniform, and decays within the skin-depth of the conducting surface.  As a result, only the nuclei within the skin depth (typically several microns) are excited by the NMR pulses and contribute to the signal. The optimal pulse length for a single crystal is determined by maximizing the signal intensity as a function of pulse length, but since $H_1$ is non-uniform not all of the nuclei within the skin depth are rotated by the same angle during this pulse, and thus the signal intensity varies only weakly with pulse length. In a topological insulator, protected surface states are expected to carry spin currents, which would give rise to discontinuities in the magnetization, $\Delta M$, at the surface of the material.\cite{RevModPhys.82.3045}  If this discontinuity depends on the field orientation relative to the normal of the plane of crystallites (along (001)), then it may explain the strong angular variation we observe in the intensity of the NMR response.  For the single crystal studies, the optimal pulse length varies only minimally as a function of field orientation because  $H_1$  remains inhomogeneous, regardless of the angular dependence of the magnetization discontinuity, $\Delta M(\theta)$.  However, if the particle size is on the order of the skin depth, then  $H_1$  will remain fairly homogeneous over the sample volume, and thus the signal intensity will depend strongly on the pulse length.  If $\Delta M(\theta)$ varies strongly with $\theta$, then the optimal pulse length will also vary strongly with orientation of the crystallite.  As a result, for a random powder or ensemble of nanoparticles with a distribution of grain sizes on the order of the skin depth, the optimal pulse length can preferentially weight one orientation more than another. If $\Delta M(\theta)$ arises from surface spin currents in our Bi$_2$Se$_3$ samples, it may explain the anomalous spectra we observe.  In order to explain the suppression observed in Fig. \ref{scry_angdep_area}, we would estimate $\Delta M(90^{\circ}) \sim 0.1 \Delta M(0^{\circ})$.  Unfortunately a complete theory of the magnetic response of topological insulators remains lacking at present, however, a large paramagnetic susceptibility associated with the surface in this material has been reported recently.\cite{Zhao2014} This anomalous magnetic response of the surface may be related to the unusual angular dependence we observe in our NMR spectra, and suggests that further experiments are needed to clarify the behavior of these materials.

In modeling the spectra of the powder and nanoparticles, we have assumed that the electronic behavior is uniform for all grains and crystallites.  The fact that the powder spectra from ground single crystals and the nano-particle spectra exhibit similar features suggests that such an approach may be justified, but we can not rule out the possibility that the electronic response in the distribution of small grains differs from the bulk single crystal.  This difference may be manifest as different relaxation rates, $T_2^{-1}$, which may, in turn, vary strongly with orientation and/or crystallite dimensions and therefore explain the discrepancy observed in Fig. \ref{powder_simspec}.

In summary, we have measured the $^{209}$Bi NMR response in nanoparticles, ground powders, and single crystals of Bi$_2$Se$_3$, and have observed an anomalous suppression of signal at high angles of the field relative to the (001) direction.  Our experiments rule out either  a partial alignment of the crystallites by the magnetic field, or an anomalous orientational dependence of the spin-spin decoherence time, $T_2^{-1}$.  We suggest that our observations may be related to an anomalous magnetic response arising from the surface currents associated with the topologically protected states in this material.   Further experiments are required to fully characterize the response of these materials to NMR pulses.

\begin{acknowledgments}

We  thank N. apRoberts-Warren and P. Klavins for the construction of the Bridgman furnace, and A. Guaghan and  R. Singh for stimulating discussions. This work was supported by the National Science Foundation (Grant No.\ DMR-1005393), the Alfred H. and Marie E. Gibeling Fellowship, and the University of California, Davis Graduate Research Mentorship.

\end{acknowledgments}

%\bibliography{Bi2Se3_pwdr_NMR_v6}
\bibliography{Bi2Se3_pwdr_NMR_v6}

%merlin.mbs apsrev4-1.bst 2010-07-25 4.21a (PWD, AO, DPC) hacked
%Control: key (0)
%Control: author (8) initials jnrlst
%Control: editor formatted (1) identically to author
%Control: production of article title (-1) disabled
%Control: page (0) single
%Control: year (1) truncated
%Control: production of eprint (0) enabled
\begin{thebibliography}{23}%
\makeatletter
\providecommand \@ifxundefined [1]{%
 \@ifx{#1\undefined}
}%
\providecommand \@ifnum [1]{%
 \ifnum #1\expandafter \@firstoftwo
 \else \expandafter \@secondoftwo
 \fi
}%
\providecommand \@ifx [1]{%
 \ifx #1\expandafter \@firstoftwo
 \else \expandafter \@secondoftwo
 \fi
}%
\providecommand \natexlab [1]{#1}%
\providecommand \enquote  [1]{``#1''}%
\providecommand \bibnamefont  [1]{#1}%
\providecommand \bibfnamefont [1]{#1}%
\providecommand \citenamefont [1]{#1}%
\providecommand \href@noop [0]{\@secondoftwo}%
\providecommand \href [0]{\begingroup \@sanitize@url \@href}%
\providecommand \@href[1]{\@@startlink{#1}\@@href}%
\providecommand \@@href[1]{\endgroup#1\@@endlink}%
\providecommand \@sanitize@url [0]{\catcode `\\12\catcode `\$12\catcode
  `\&12\catcode `\#12\catcode `\^12\catcode `\_12\catcode `\%12\relax}%
\providecommand \@@startlink[1]{}%
\providecommand \@@endlink[0]{}%
\providecommand \url  [0]{\begingroup\@sanitize@url \@url }%
\providecommand \@url [1]{\endgroup\@href {#1}{\urlprefix }}%
\providecommand \urlprefix  [0]{URL }%
\providecommand \Eprint [0]{\href }%
\providecommand \doibase [0]{http://dx.doi.org/}%
\providecommand \selectlanguage [0]{\@gobble}%
\providecommand \bibinfo  [0]{\@secondoftwo}%
\providecommand \bibfield  [0]{\@secondoftwo}%
\providecommand \translation [1]{[#1]}%
\providecommand \BibitemOpen [0]{}%
\providecommand \bibitemStop [0]{}%
\providecommand \bibitemNoStop [0]{.\EOS\space}%
\providecommand \EOS [0]{\spacefactor3000\relax}%
\providecommand \BibitemShut  [1]{\csname bibitem#1\endcsname}%
\let\auto@bib@innerbib\@empty
%</preamble>
\bibitem [{\citenamefont {Pesin}\ and\ \citenamefont
  {MacDonald}(2012)}]{Pesin2012}%
  \BibitemOpen
  \bibfield  {author} {\bibinfo {author} {\bibfnamefont {D.}~\bibnamefont
  {Pesin}}\ and\ \bibinfo {author} {\bibfnamefont {A.~H.}\ \bibnamefont
  {MacDonald}},\ }\href {http://dx.doi.org/10.1038/nmat3305} {\bibfield
  {journal} {\bibinfo  {journal} {Nat. Mater.}\ }\textbf {\bibinfo {volume}
  {11}},\ \bibinfo {pages} {409} (\bibinfo {year} {2012})}\BibitemShut
  {NoStop}%
\bibitem [{\citenamefont {Chen}\ \emph {et~al.}(2009)\citenamefont {Chen},
  \citenamefont {Analytis}, \citenamefont {Chu}, \citenamefont {Liu},
  \citenamefont {Mo}, \citenamefont {Qi}, \citenamefont {Zhang}, \citenamefont
  {Lu}, \citenamefont {Dai}, \citenamefont {Fang}, \citenamefont {Zhang},
  \citenamefont {Fisher}, \citenamefont {Hussain},\ and\ \citenamefont
  {Shen}}]{Chen10072009}%
  \BibitemOpen
  \bibfield  {author} {\bibinfo {author} {\bibfnamefont {Y.~L.}\ \bibnamefont
  {Chen}}, \bibinfo {author} {\bibfnamefont {J.~G.}\ \bibnamefont {Analytis}},
  \bibinfo {author} {\bibfnamefont {J.-H.}\ \bibnamefont {Chu}}, \bibinfo
  {author} {\bibfnamefont {Z.~K.}\ \bibnamefont {Liu}}, \bibinfo {author}
  {\bibfnamefont {S.-K.}\ \bibnamefont {Mo}}, \bibinfo {author} {\bibfnamefont
  {X.~L.}\ \bibnamefont {Qi}}, \bibinfo {author} {\bibfnamefont {H.~J.}\
  \bibnamefont {Zhang}}, \bibinfo {author} {\bibfnamefont {D.~H.}\ \bibnamefont
  {Lu}}, \bibinfo {author} {\bibfnamefont {X.}~\bibnamefont {Dai}}, \bibinfo
  {author} {\bibfnamefont {Z.}~\bibnamefont {Fang}}, \bibinfo {author}
  {\bibfnamefont {S.~C.}\ \bibnamefont {Zhang}}, \bibinfo {author}
  {\bibfnamefont {I.~R.}\ \bibnamefont {Fisher}}, \bibinfo {author}
  {\bibfnamefont {Z.}~\bibnamefont {Hussain}}, \ and\ \bibinfo {author}
  {\bibfnamefont {Z.-X.}\ \bibnamefont {Shen}},\ }\href {\doibase
  10.1126/science.1173034} {\bibfield  {journal} {\bibinfo  {journal}
  {Science}\ }\textbf {\bibinfo {volume} {325}},\ \bibinfo {pages} {178}
  (\bibinfo {year} {2009})}\BibitemShut {NoStop}%
\bibitem [{\citenamefont {Liu}\ \emph {et~al.}(2009)\citenamefont {Liu},
  \citenamefont {Liu}, \citenamefont {Xu}, \citenamefont {Qi},\ and\
  \citenamefont {Zhang}}]{PhysRevLett.102.156603}%
  \BibitemOpen
  \bibfield  {author} {\bibinfo {author} {\bibfnamefont {Q.}~\bibnamefont
  {Liu}}, \bibinfo {author} {\bibfnamefont {C.-X.}\ \bibnamefont {Liu}},
  \bibinfo {author} {\bibfnamefont {C.}~\bibnamefont {Xu}}, \bibinfo {author}
  {\bibfnamefont {X.-L.}\ \bibnamefont {Qi}}, \ and\ \bibinfo {author}
  {\bibfnamefont {S.-C.}\ \bibnamefont {Zhang}},\ }\href {\doibase
  10.1103/PhysRevLett.102.156603} {\bibfield  {journal} {\bibinfo  {journal}
  {Phys. Rev. Lett.}\ }\textbf {\bibinfo {volume} {102}},\ \bibinfo {pages}
  {156603} (\bibinfo {year} {2009})}\BibitemShut {NoStop}%
\bibitem [{\citenamefont {Ando}(2013)}]{JPSJ.82.102001}%
  \BibitemOpen
  \bibfield  {author} {\bibinfo {author} {\bibfnamefont {Y.}~\bibnamefont
  {Ando}},\ }\href {\doibase 10.1143/JPSJ.82.102001} {\bibfield  {journal}
  {\bibinfo  {journal} {J. Phys. Soc. Jpn.}\ }\textbf {\bibinfo {volume}
  {82}},\ \bibinfo {pages} {102001} (\bibinfo {year} {2013})}\BibitemShut
  {NoStop}%
\bibitem [{\citenamefont {Moore}(2010)}]{Moore2010}%
  \BibitemOpen
  \bibfield  {author} {\bibinfo {author} {\bibfnamefont {J.~E.}\ \bibnamefont
  {Moore}},\ }\href {http://dx.doi.org/10.1038/nature08916} {\bibfield
  {journal} {\bibinfo  {journal} {Nature}\ }\textbf {\bibinfo {volume} {464}},\
  \bibinfo {pages} {194} (\bibinfo {year} {2010})}\BibitemShut {NoStop}%
\bibitem [{\citenamefont {Xia}\ \emph {et~al.}(2009)\citenamefont {Xia},
  \citenamefont {Qian}, \citenamefont {Hsieh}, \citenamefont {Wray},
  \citenamefont {Pal}, \citenamefont {Lin}, \citenamefont {Bansil},
  \citenamefont {Grauer}, \citenamefont {Hor}, \citenamefont {Cava},\ and\
  \citenamefont {Hasan}}]{Xia2009}%
  \BibitemOpen
  \bibfield  {author} {\bibinfo {author} {\bibfnamefont {Y.}~\bibnamefont
  {Xia}}, \bibinfo {author} {\bibfnamefont {D.}~\bibnamefont {Qian}}, \bibinfo
  {author} {\bibfnamefont {D.}~\bibnamefont {Hsieh}}, \bibinfo {author}
  {\bibfnamefont {L.}~\bibnamefont {Wray}}, \bibinfo {author} {\bibfnamefont
  {A.}~\bibnamefont {Pal}}, \bibinfo {author} {\bibfnamefont {H.}~\bibnamefont
  {Lin}}, \bibinfo {author} {\bibfnamefont {A.}~\bibnamefont {Bansil}},
  \bibinfo {author} {\bibfnamefont {D.}~\bibnamefont {Grauer}}, \bibinfo
  {author} {\bibfnamefont {Y.~S.}\ \bibnamefont {Hor}}, \bibinfo {author}
  {\bibfnamefont {R.~J.}\ \bibnamefont {Cava}}, \ and\ \bibinfo {author}
  {\bibfnamefont {M.~Z.}\ \bibnamefont {Hasan}},\ }\href
  {http://dx.doi.org/10.1038/nphys1274} {\bibfield  {journal} {\bibinfo
  {journal} {Nat Phys}\ }\textbf {\bibinfo {volume} {5}},\ \bibinfo {pages}
  {398} (\bibinfo {year} {2009})}\BibitemShut {NoStop}%
\bibitem [{\citenamefont {Nisson}\ \emph {et~al.}(2013)\citenamefont {Nisson},
  \citenamefont {Dioguardi}, \citenamefont {Klavins}, \citenamefont {Lin},
  \citenamefont {Shirer}, \citenamefont {Shockley}, \citenamefont {Crocker},\
  and\ \citenamefont {Curro}}]{PhysRevB.87.195202}%
  \BibitemOpen
  \bibfield  {author} {\bibinfo {author} {\bibfnamefont {D.~M.}\ \bibnamefont
  {Nisson}}, \bibinfo {author} {\bibfnamefont {A.~P.}\ \bibnamefont
  {Dioguardi}}, \bibinfo {author} {\bibfnamefont {P.}~\bibnamefont {Klavins}},
  \bibinfo {author} {\bibfnamefont {C.~H.}\ \bibnamefont {Lin}}, \bibinfo
  {author} {\bibfnamefont {K.}~\bibnamefont {Shirer}}, \bibinfo {author}
  {\bibfnamefont {A.~C.}\ \bibnamefont {Shockley}}, \bibinfo {author}
  {\bibfnamefont {J.}~\bibnamefont {Crocker}}, \ and\ \bibinfo {author}
  {\bibfnamefont {N.~J.}\ \bibnamefont {Curro}},\ }\href {\doibase
  10.1103/PhysRevB.87.195202} {\bibfield  {journal} {\bibinfo  {journal} {Phys.
  Rev. B}\ }\textbf {\bibinfo {volume} {87}},\ \bibinfo {pages} {195202}
  (\bibinfo {year} {2013})}\BibitemShut {NoStop}%
\bibitem [{\citenamefont {Hyde}\ \emph {et~al.}(1974)\citenamefont {Hyde},
  \citenamefont {Beale}, \citenamefont {Spain},\ and\ \citenamefont
  {Woollam}}]{hyde1974}%
  \BibitemOpen
  \bibfield  {author} {\bibinfo {author} {\bibfnamefont {G.}~\bibnamefont
  {Hyde}}, \bibinfo {author} {\bibfnamefont {H.}~\bibnamefont {Beale}},
  \bibinfo {author} {\bibfnamefont {I.}~\bibnamefont {Spain}}, \ and\ \bibinfo
  {author} {\bibfnamefont {J.}~\bibnamefont {Woollam}},\ }\href {\doibase
  10.1016/S0022-3697(74)80186-1} {\bibfield  {journal} {\bibinfo  {journal} {J.
  Phys. Chem. Solids}\ }\textbf {\bibinfo {volume} {35}},\ \bibinfo {pages}
  {1719 } (\bibinfo {year} {1974})}\BibitemShut {NoStop}%
\bibitem [{\citenamefont {Butch}\ \emph {et~al.}(2010)\citenamefont {Butch},
  \citenamefont {Kirshenbaum}, \citenamefont {Syers}, \citenamefont {Sushkov},
  \citenamefont {Jenkins}, \citenamefont {Drew},\ and\ \citenamefont
  {Paglione}}]{butch2010}%
  \BibitemOpen
  \bibfield  {author} {\bibinfo {author} {\bibfnamefont {N.~P.}\ \bibnamefont
  {Butch}}, \bibinfo {author} {\bibfnamefont {K.}~\bibnamefont {Kirshenbaum}},
  \bibinfo {author} {\bibfnamefont {P.}~\bibnamefont {Syers}}, \bibinfo
  {author} {\bibfnamefont {A.~B.}\ \bibnamefont {Sushkov}}, \bibinfo {author}
  {\bibfnamefont {G.~S.}\ \bibnamefont {Jenkins}}, \bibinfo {author}
  {\bibfnamefont {H.~D.}\ \bibnamefont {Drew}}, \ and\ \bibinfo {author}
  {\bibfnamefont {J.}~\bibnamefont {Paglione}},\ }\href {\doibase
  10.1103/PhysRevB.81.241301} {\bibfield  {journal} {\bibinfo  {journal} {Phys.
  Rev. B}\ }\textbf {\bibinfo {volume} {81}},\ \bibinfo {pages} {241301}
  (\bibinfo {year} {2010})}\BibitemShut {NoStop}%
\bibitem [{\citenamefont {Hasan}\ and\ \citenamefont
  {Kane}(2010)}]{RevModPhys.82.3045}%
  \BibitemOpen
  \bibfield  {author} {\bibinfo {author} {\bibfnamefont {M.~Z.}\ \bibnamefont
  {Hasan}}\ and\ \bibinfo {author} {\bibfnamefont {C.~L.}\ \bibnamefont
  {Kane}},\ }\href {\doibase 10.1103/RevModPhys.82.3045} {\bibfield  {journal}
  {\bibinfo  {journal} {Rev. Mod. Phys.}\ }\textbf {\bibinfo {volume} {82}},\
  \bibinfo {pages} {3045} (\bibinfo {year} {2010})}\BibitemShut {NoStop}%
\bibitem [{\citenamefont {Zhang}\ \emph {et~al.}(2009)\citenamefont {Zhang},
  \citenamefont {Liu}, \citenamefont {Qi}, \citenamefont {Dai}, \citenamefont
  {Fang},\ and\ \citenamefont {Zhang}}]{Zhang2009}%
  \BibitemOpen
  \bibfield  {author} {\bibinfo {author} {\bibfnamefont {H.}~\bibnamefont
  {Zhang}}, \bibinfo {author} {\bibfnamefont {C.-X.}\ \bibnamefont {Liu}},
  \bibinfo {author} {\bibfnamefont {X.-L.}\ \bibnamefont {Qi}}, \bibinfo
  {author} {\bibfnamefont {X.}~\bibnamefont {Dai}}, \bibinfo {author}
  {\bibfnamefont {Z.}~\bibnamefont {Fang}}, \ and\ \bibinfo {author}
  {\bibfnamefont {S.-C.}\ \bibnamefont {Zhang}},\ }\href
  {http://dx.doi.org/10.1038/nphys1270} {\bibfield  {journal} {\bibinfo
  {journal} {Nat Phys}\ }\textbf {\bibinfo {volume} {5}},\ \bibinfo {pages}
  {438} (\bibinfo {year} {2009})}\BibitemShut {NoStop}%
\bibitem [{\citenamefont {Bianchi}\ \emph {et~al.}(2010)\citenamefont
  {Bianchi}, \citenamefont {Guan}, \citenamefont {Bao}, \citenamefont {Mi},
  \citenamefont {Iversen}, \citenamefont {King},\ and\ \citenamefont
  {Hofmann}}]{Bianchi2010}%
  \BibitemOpen
  \bibfield  {author} {\bibinfo {author} {\bibfnamefont {M.}~\bibnamefont
  {Bianchi}}, \bibinfo {author} {\bibfnamefont {D.}~\bibnamefont {Guan}},
  \bibinfo {author} {\bibfnamefont {S.}~\bibnamefont {Bao}}, \bibinfo {author}
  {\bibfnamefont {J.}~\bibnamefont {Mi}}, \bibinfo {author} {\bibfnamefont
  {B.~B.}\ \bibnamefont {Iversen}}, \bibinfo {author} {\bibfnamefont
  {P.~D.~C.}\ \bibnamefont {King}}, \ and\ \bibinfo {author} {\bibfnamefont
  {P.}~\bibnamefont {Hofmann}},\ }\href {http://dx.doi.org/10.1038/ncomms1131}
  {\bibfield  {journal} {\bibinfo  {journal} {Nat Commun}\ }\textbf {\bibinfo
  {volume} {1}},\ \bibinfo {pages} {128} (\bibinfo {year} {2010})}\BibitemShut
  {NoStop}%
\bibitem [{\citenamefont {Ren}\ \emph {et~al.}(2012)\citenamefont {Ren},
  \citenamefont {Taskin}, \citenamefont {Sasaki}, \citenamefont {Segawa},\ and\
  \citenamefont {Ando}}]{PhysRevB.85.155301}%
  \BibitemOpen
  \bibfield  {author} {\bibinfo {author} {\bibfnamefont {Z.}~\bibnamefont
  {Ren}}, \bibinfo {author} {\bibfnamefont {A.~A.}\ \bibnamefont {Taskin}},
  \bibinfo {author} {\bibfnamefont {S.}~\bibnamefont {Sasaki}}, \bibinfo
  {author} {\bibfnamefont {K.}~\bibnamefont {Segawa}}, \ and\ \bibinfo {author}
  {\bibfnamefont {Y.}~\bibnamefont {Ando}},\ }\href {\doibase
  10.1103/PhysRevB.85.155301} {\bibfield  {journal} {\bibinfo  {journal} {Phys.
  Rev. B}\ }\textbf {\bibinfo {volume} {85}},\ \bibinfo {pages} {155301}
  (\bibinfo {year} {2012})}\BibitemShut {NoStop}%
\bibitem [{\citenamefont {Vazifeh}\ and\ \citenamefont
  {Franz}(2012)}]{PhysRevB.86.045451}%
  \BibitemOpen
  \bibfield  {author} {\bibinfo {author} {\bibfnamefont {M.~M.}\ \bibnamefont
  {Vazifeh}}\ and\ \bibinfo {author} {\bibfnamefont {M.}~\bibnamefont
  {Franz}},\ }\href {\doibase 10.1103/PhysRevB.86.045451} {\bibfield  {journal}
  {\bibinfo  {journal} {Phys. Rev. B}\ }\textbf {\bibinfo {volume} {86}},\
  \bibinfo {pages} {045451} (\bibinfo {year} {2012})}\BibitemShut {NoStop}%
\bibitem [{\citenamefont {Koumoulis}\ \emph {et~al.}(2013)\citenamefont
  {Koumoulis}, \citenamefont {Chasapis}, \citenamefont {Taylor}, \citenamefont
  {Lake}, \citenamefont {King}, \citenamefont {Jarenwattananon}, \citenamefont
  {Fiete}, \citenamefont {Kanatzidis},\ and\ \citenamefont
  {Bouchard}}]{PhysRevLett.110.026602}%
  \BibitemOpen
  \bibfield  {author} {\bibinfo {author} {\bibfnamefont {D.}~\bibnamefont
  {Koumoulis}}, \bibinfo {author} {\bibfnamefont {T.~C.}\ \bibnamefont
  {Chasapis}}, \bibinfo {author} {\bibfnamefont {R.~E.}\ \bibnamefont
  {Taylor}}, \bibinfo {author} {\bibfnamefont {M.~P.}\ \bibnamefont {Lake}},
  \bibinfo {author} {\bibfnamefont {D.}~\bibnamefont {King}}, \bibinfo {author}
  {\bibfnamefont {N.~N.}\ \bibnamefont {Jarenwattananon}}, \bibinfo {author}
  {\bibfnamefont {G.~A.}\ \bibnamefont {Fiete}}, \bibinfo {author}
  {\bibfnamefont {M.~G.}\ \bibnamefont {Kanatzidis}}, \ and\ \bibinfo {author}
  {\bibfnamefont {L.-S.}\ \bibnamefont {Bouchard}},\ }\href {\doibase
  10.1103/PhysRevLett.110.026602} {\bibfield  {journal} {\bibinfo  {journal}
  {Phys. Rev. Lett.}\ }\textbf {\bibinfo {volume} {110}},\ \bibinfo {pages}
  {026602} (\bibinfo {year} {2013})}\BibitemShut {NoStop}%
\bibitem [{\citenamefont {Lind}\ \emph {et~al.}(2005)\citenamefont {Lind},
  \citenamefont {Lidin},\ and\ \citenamefont
  {H\"aussermann}}]{LatticeConstants1998}%
  \BibitemOpen
  \bibfield  {author} {\bibinfo {author} {\bibfnamefont {H.}~\bibnamefont
  {Lind}}, \bibinfo {author} {\bibfnamefont {S.}~\bibnamefont {Lidin}}, \ and\
  \bibinfo {author} {\bibfnamefont {U.}~\bibnamefont {H\"aussermann}},\ }\href
  {\doibase 10.1103/PhysRevB.72.184101} {\bibfield  {journal} {\bibinfo
  {journal} {Phys. Rev. B}\ }\textbf {\bibinfo {volume} {72}},\ \bibinfo
  {pages} {184101} (\bibinfo {year} {2005})}\BibitemShut {NoStop}%
\bibitem [{\citenamefont {Kong}\ \emph {et~al.}(2010)\citenamefont {Kong},
  \citenamefont {Randel}, \citenamefont {Peng}, \citenamefont {Cha},
  \citenamefont {Meister}, \citenamefont {Lai}, \citenamefont {Chen},
  \citenamefont {Shen}, \citenamefont {Manoharan},\ and\ \citenamefont
  {Cui}}]{Kong2010}%
  \BibitemOpen
  \bibfield  {author} {\bibinfo {author} {\bibfnamefont {D.}~\bibnamefont
  {Kong}}, \bibinfo {author} {\bibfnamefont {J.~C.}\ \bibnamefont {Randel}},
  \bibinfo {author} {\bibfnamefont {H.}~\bibnamefont {Peng}}, \bibinfo {author}
  {\bibfnamefont {J.~J.}\ \bibnamefont {Cha}}, \bibinfo {author} {\bibfnamefont
  {S.}~\bibnamefont {Meister}}, \bibinfo {author} {\bibfnamefont
  {K.}~\bibnamefont {Lai}}, \bibinfo {author} {\bibfnamefont {Y.}~\bibnamefont
  {Chen}}, \bibinfo {author} {\bibfnamefont {Z.-X.}\ \bibnamefont {Shen}},
  \bibinfo {author} {\bibfnamefont {H.~C.}\ \bibnamefont {Manoharan}}, \ and\
  \bibinfo {author} {\bibfnamefont {Y.}~\bibnamefont {Cui}},\ }\href {\doibase
  10.1021/nl903663a} {\bibfield  {journal} {\bibinfo  {journal} {Nano Lett.}\
  }\textbf {\bibinfo {volume} {10}},\ \bibinfo {pages} {329} (\bibinfo {year}
  {2010})}\BibitemShut {NoStop}%
\bibitem [{\citenamefont {Slichter}(1992)}]{CPSbook}%
  \BibitemOpen
  \bibfield  {author} {\bibinfo {author} {\bibfnamefont {C.~P.}\ \bibnamefont
  {Slichter}},\ }\href@noop {} {\emph {\bibinfo {title} {Principles of Nuclear
  Magnetic Resonance}}},\ \bibinfo {edition} {3rd}\ ed.\ (\bibinfo  {publisher}
  {Springer-Verlag},\ \bibinfo {year} {1992})\BibitemShut {NoStop}%
\bibitem [{\citenamefont {Young}\ \emph {et~al.}(2012)\citenamefont {Young},
  \citenamefont {Lai}, \citenamefont {Xu}, \citenamefont {Yang}, \citenamefont
  {Gu}, \citenamefont {Pan}, \citenamefont {Valla}, \citenamefont {Shu},
  \citenamefont {Sankar},\ and\ \citenamefont {Chou}}]{PhysRevB.86.075137}%
  \BibitemOpen
  \bibfield  {author} {\bibinfo {author} {\bibfnamefont {B.-L.}\ \bibnamefont
  {Young}}, \bibinfo {author} {\bibfnamefont {Z.-Y.}\ \bibnamefont {Lai}},
  \bibinfo {author} {\bibfnamefont {Z.}~\bibnamefont {Xu}}, \bibinfo {author}
  {\bibfnamefont {A.}~\bibnamefont {Yang}}, \bibinfo {author} {\bibfnamefont
  {G.~D.}\ \bibnamefont {Gu}}, \bibinfo {author} {\bibfnamefont {Z.-H.}\
  \bibnamefont {Pan}}, \bibinfo {author} {\bibfnamefont {T.}~\bibnamefont
  {Valla}}, \bibinfo {author} {\bibfnamefont {G.~J.}\ \bibnamefont {Shu}},
  \bibinfo {author} {\bibfnamefont {R.}~\bibnamefont {Sankar}}, \ and\ \bibinfo
  {author} {\bibfnamefont {F.~C.}\ \bibnamefont {Chou}},\ }\href {\doibase
  10.1103/PhysRevB.86.075137} {\bibfield  {journal} {\bibinfo  {journal} {Phys.
  Rev. B}\ }\textbf {\bibinfo {volume} {86}},\ \bibinfo {pages} {075137}
  (\bibinfo {year} {2012})}\BibitemShut {NoStop}%
\bibitem [{\citenamefont {Reyes}\ \emph {et~al.}(1994)\citenamefont {Reyes},
  \citenamefont {Heffner}, \citenamefont {Canfield}, \citenamefont {Thompson},\
  and\ \citenamefont {Fisk}}]{PhysRevB.49.16321}%
  \BibitemOpen
  \bibfield  {author} {\bibinfo {author} {\bibfnamefont {A.~P.}\ \bibnamefont
  {Reyes}}, \bibinfo {author} {\bibfnamefont {R.~H.}\ \bibnamefont {Heffner}},
  \bibinfo {author} {\bibfnamefont {P.~C.}\ \bibnamefont {Canfield}}, \bibinfo
  {author} {\bibfnamefont {J.~D.}\ \bibnamefont {Thompson}}, \ and\ \bibinfo
  {author} {\bibfnamefont {Z.}~\bibnamefont {Fisk}},\ }\href {\doibase
  10.1103/PhysRevB.49.16321} {\bibfield  {journal} {\bibinfo  {journal} {Phys.
  Rev. B}\ }\textbf {\bibinfo {volume} {49}},\ \bibinfo {pages} {16321}
  (\bibinfo {year} {1994})}\BibitemShut {NoStop}%
\bibitem [{\citenamefont {Curro}\ and\ \citenamefont
  {Slichter}(1998)}]{NickCharlieEFK}%
  \BibitemOpen
  \bibfield  {author} {\bibinfo {author} {\bibfnamefont {N.~J.}\ \bibnamefont
  {Curro}}\ and\ \bibinfo {author} {\bibfnamefont {C.~P.}\ \bibnamefont
  {Slichter}},\ }\href {\doibase http://dx.doi.org/10.1006/jmre.1997.1295}
  {\bibfield  {journal} {\bibinfo  {journal} {J. Mag. Res.}\ }\textbf {\bibinfo
  {volume} {130}},\ \bibinfo {pages} {186 } (\bibinfo {year}
  {1998})}\BibitemShut {NoStop}%
\bibitem [{\citenamefont {Bush}\ \emph {et~al.}(2003)\citenamefont {Bush},
  \citenamefont {Gippius}, \citenamefont {Zalesskii},\ and\ \citenamefont
  {Morozova}}]{BiFeO3NMR}%
  \BibitemOpen
  \bibfield  {author} {\bibinfo {author} {\bibfnamefont {A.}~\bibnamefont
  {Bush}}, \bibinfo {author} {\bibfnamefont {A.}~\bibnamefont {Gippius}},
  \bibinfo {author} {\bibfnamefont {A.}~\bibnamefont {Zalesskii}}, \ and\
  \bibinfo {author} {\bibfnamefont {E.}~\bibnamefont {Morozova}},\ }\href
  {\doibase 10.1134/1.1630133} {\bibfield  {journal} {\bibinfo  {journal} {JETP
  Lett.}\ }\textbf {\bibinfo {volume} {78}},\ \bibinfo {pages} {389} (\bibinfo
  {year} {2003})}\BibitemShut {NoStop}%
\bibitem [{\citenamefont {Zhao}\ \emph {et~al.}(2014)\citenamefont {Zhao},
  \citenamefont {Deng}, \citenamefont {Korzhovska}, \citenamefont {Chen},
  \citenamefont {Konczykowski}, \citenamefont {Hruban}, \citenamefont
  {Oganesyan},\ and\ \citenamefont {Krusin-Elbaum}}]{Zhao2014}%
  \BibitemOpen
  \bibfield  {author} {\bibinfo {author} {\bibfnamefont {L.}~\bibnamefont
  {Zhao}}, \bibinfo {author} {\bibfnamefont {H.}~\bibnamefont {Deng}}, \bibinfo
  {author} {\bibfnamefont {I.}~\bibnamefont {Korzhovska}}, \bibinfo {author}
  {\bibfnamefont {Z.}~\bibnamefont {Chen}}, \bibinfo {author} {\bibfnamefont
  {M.}~\bibnamefont {Konczykowski}}, \bibinfo {author} {\bibfnamefont
  {A.}~\bibnamefont {Hruban}}, \bibinfo {author} {\bibfnamefont
  {V.}~\bibnamefont {Oganesyan}}, \ and\ \bibinfo {author} {\bibfnamefont
  {L.}~\bibnamefont {Krusin-Elbaum}},\ }\href {\doibase 10.1038/nmat3962}
  {\bibfield  {journal} {\bibinfo  {journal} {Nat. Mater.}\ }\textbf {\bibinfo
  {volume} {13}},\ \bibinfo {pages} {580 } (\bibinfo {year}
  {2014})}\BibitemShut {NoStop}%
\end{thebibliography}%
\end{document}